\documentclass[%
 aip,
rsi,%
 amsmath,amssymb,
 reprint,%
]{revtex4-1}

\usepackage{graphicx}% Include figure files
\usepackage{dcolumn}% Align table columns on decimal point
\usepackage{bm}% bold math
\begin{document}

\preprint{AIP/123-QED}

\title{Quantitative analysis of the interaction between a dc SQUID and an integrated micromechanical doubly clamped cantilever}% Force line breaks with \\

\author{Majdi Salman}
\altaffiliation{Electronic mail: Salmanm2@cardiff.ac.uk}%Lines break automatically or can be forced with \\
\author{Georgina M Klemencic}
\author{Soumen Mandal}
\author{Scott Manifold}
 \affiliation{School of Physics and Astronomy, Cardiff University, Queen's Building, The Parade, 
 \\ Cardiff, CF24 3AA, United Kingdom}%
 
 \author{Luqman Mustafa}
 \affiliation{Centre for Innovation Competence SiLi-nano, Martin-Luther-University Halle-Wittenberg, 
 Karl-Freiherr-von-Fritsch-Strasse 3, 06120 Halle (Saale), Germany }%

\author{Oliver A Williams}
\author{Sean R Giblin}
\affiliation{School of Physics and Astronomy, Cardiff University, Queen's Building, The Parade, 
 \\ Cardiff, CF24 3AA, United Kingdom}%
 
\date{\today}% It is always \today, today,
             %  but any date may be explicitly specified

\begin{abstract}
Based on the superconducting quantum interference device (SQUID) equations described by the resistively- and capacitively-shunted junction model coupled to the equation of motion of a damped harmonic oscillator, we provide simulations to quantitatively describe the interaction between a dc SQUID and an integrated doubly clamped cantilever. We have chosen to investigate an existing experimental configuration and have explored the motion of the cantilever and the reaction of the SQUID as a function of the voltage-flux $V(\Phi)$ characteristics. We clearly observe the Lorentz force back-action interaction and demonstrate how a sharp transition state drives the system into a nonlinear-like regime, and modulates the cantilever displacement amplitude, simply by tuning the SQUID parameters.
\end{abstract}

\pacs{Valid PACS appear here}% PACS, the Physics and Astronomy
                             % Classification Scheme.
% \keywords{Suggested keywords}%Use showkeys class option if keyword
                              %display desired
\maketitle

\section{\label{sec:level1} Introduction}

Theoretical and experimental studies\cite{Etaki,Buks, Armour, Fazio, DArmour, Ventra, Ella} of linear and nonlinear micro and nanomechanical resonators are of great interest as they can be used for sensitive force and displacement measurements. The physical parameters of the resonators can also be tuned to observe the transition from the classical to quantum regimes with relative experimental ease, enabling observations of macroscopic quantum systems.\cite{Schwab1}~Significant experimental progress in the detection of resonators as they enter the quantum ground state has been achieved by capactiave coupling to superconducting flux qubits,\cite{Connell} and quantum state control of a mechanical drum resonator in a superconducting resonant circuit has been achieved by phonon-photon coupling.\cite{Teufel, CSchwab}~The state detection is an integral part of any coupled resonator system as the coupling mechanism is implicit in any experimental endeavour.

Considering a doubly clamped cantilever, it is obvious that as the cantilever oscillates the displacement changes, and the transduction technique will cause a back-action that influences the cantilever position.\cite{Caves} ~The impact of back-action can be positive in terms of cooling\cite{DArmour} and squeezing the resonator motion,\cite{ Almog, Liu,Zoller} and coupling and synchronising multiple resonators.\cite{Wiesenfeld, Rogers}~Depending on the specific transduction technique, back-action can be due to radiation pressure,\cite{Kippenberg} electron tunnelling,\cite{ Steele} or photothermal effects.\cite{ Jourdan} 
\begin{figure}
\centerline{\includegraphics[width=3.6 in,angle=0]{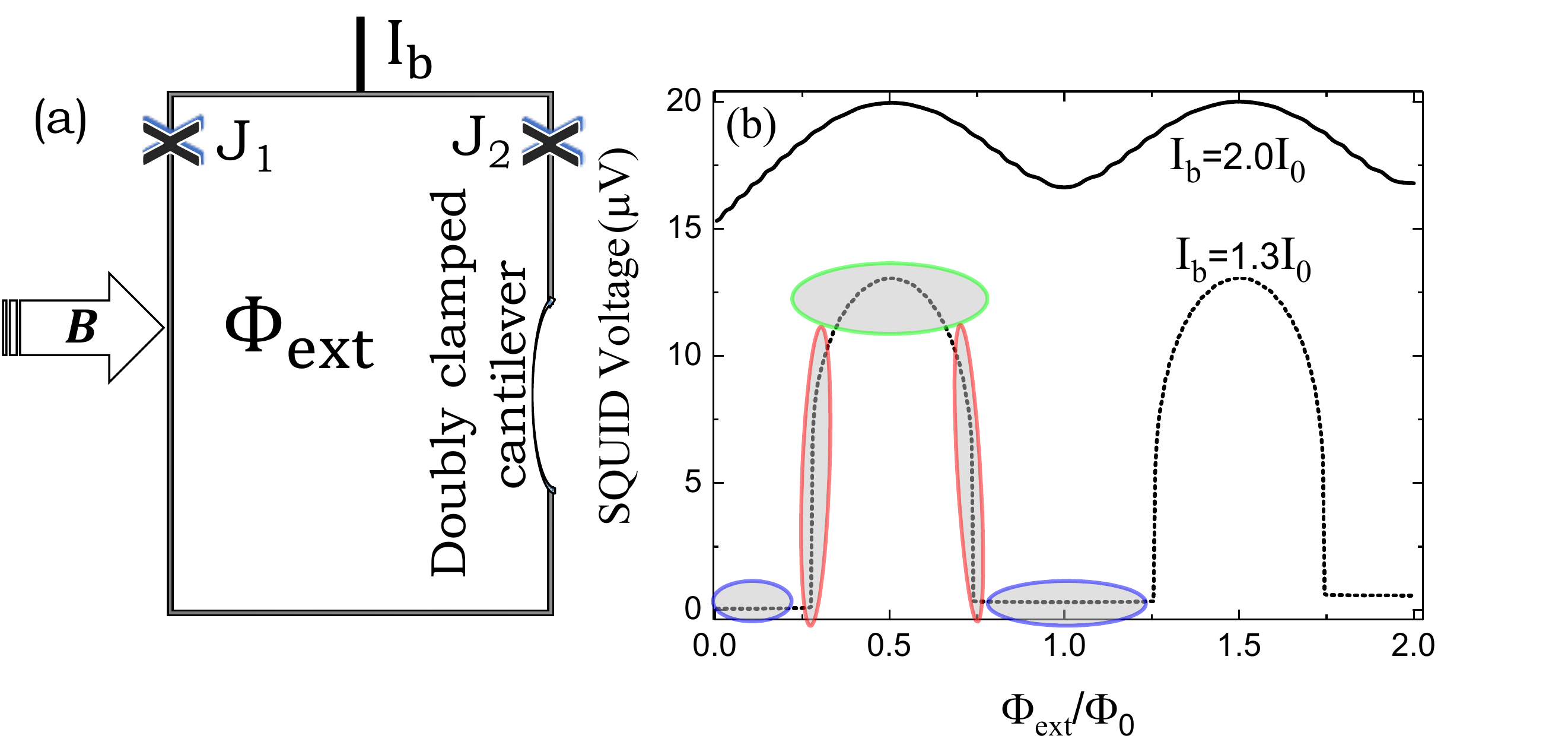}}
\caption{(a) scheme for the dc SQUID displacement detector in which the two Josephson junctions are labelled by $\mathrm {J_{1}}$ and $\mathrm {J_{2}}$. The cantilever displacement is out-of-plane,
and the applied magnetic field, $B$, is in-plane. (b) $V(\Phi)$ characteristics for a dc SQUID with $\beta_L=0.115$ and $\beta_C=1.61$. Four regimes are identified: $(i)$ the simple oscillatory regime where the bias current, $I_b=2.0I_0$. The other regimes are: $(ii)$ the rapidly changing regime (red), $(iii)$ the zero voltage response regime (blue), and $(iv)$  the intermediate regime (green).}
\label{Fig:Vphi}
\end{figure}

Previous experiments have used a dc SQUID to detect the motion of a suspended doubly clamped cantilever integrated directly into a SQUID loop,\cite{Etaki,Poot} and for a torsional SQUID cantilever.\cite{SEtaki} For this SQUID-based transduction scheme, the back-action has a simple inductive component caused by the Lorentz force due to the circulating current.\cite{SEtaki,Poot, Shevchuk}  Experimentally, the Lorentz back-action was shown to shift the mechanical cantilever resonant frequency and quality factor by $\Delta f$ and $\Delta Q$ respectively. To understand the effect of back-action on $\Delta f$ and $\Delta Q$, two transfer functions were obtained,\cite{Poot} which are coefficients for the average circulating current expanded in the terms of the cantilever displacement, $u$, and velocity, $\dot{u}$.

In previous work, however, it was not possible to obtain the velocity-dependent transfer function in the frame of the SQUID equations coupled to the equation of motion of the doubly clamped cantilever. To simplify this issue, Poot \textit{et al}\cite{Poot} modulated the flux change in the SQUID loop caused by the cantilever oscillation. Subsequently, the total flux in the SQUID loop was assumed to be a function of the externally applied flux, $\Phi_{\mathrm{ext}}$, and the modulation of the flux due to the changing area of the loop, $\Phi \rightarrow \Phi_{\mathrm{ext}}+\Phi_{\mathrm{mod}} \cos(\omega_{\mathrm{mod}}t)$. Such a modulation can describe the influence of the back-action on $\Delta f$ and $\Delta Q$ of the cantilever when the SQUID displacement detector is tuned within limited regions of the $V(\Phi)$ curve. \cite{Poot} However, a full description of the SQUID-cantilever interaction requires a comprehensive model to provide information not only about the influence of back-action in all regions of $V(\Phi)$, but also about the amplitude, width, line shape, and responsivity, $\frac{\mathrm{d}V}{\mathrm{d}u}$, which must be calculated by linking the cantilever displacement to the SQUID voltage. Thus, the need for quantitative treatments of the unscaled SQUID equations coupled explicitly to the equation of motion for the integrated beam becomes important. Though such treatments are complicated and challenging,\cite{SupEtaki} they can be performed numerically with improving computational capabilities.  

In this paper, we simulate the interaction between a dc SQUID and an embedded micromechanical doubly clamped cantilever as experimentally demonstrated by Etaki \textit{et al}~\cite{Etaki} and shown schematically in Fig.~\ref{Fig:Vphi}(a). The SQUID-cantilever interaction is analysed in different regions of the $V(\Phi)$ curve, as shown in Fig.~\ref{Fig:Vphi}(b). Within this framework, we have explored some regions of the $V(\Phi)$ curve, where the SQUID-cantilever response is apparently strongly nonlinear. Futhermore, the back-action and the subsequent response of the SQUID is linked to the cantilever displacement. The effect of changing the SQUID operating point is discussed in depth, and it is demonstrated that the SQUID itself can be used to control the cantilever response by simple modification of the controllable SQUID parameters. 

\section{\label{sec:level2} The Model}

The model we present is based on the experimental parameters of Etaki \textit{et al}~\cite{Etaki} to allow for experimental verification of the results. 
As such the inductive screening parameter, $\beta_L$, and Stewart-McCumber parameter, $\beta_C$, are selected to be $0.115$ and $1.61$ respectively. With these values for  
$\beta_L$  and $\beta_C$,  $V(\Phi)$ characteristics of an overdamped dc SQUID are shown in Fig.~\ref{Fig:Vphi}(b) to demonstrate the possible operating points of a SQUID displacement detector. The $V(\Phi)$ curves are calculated using the time-scaled SQUID equations described by the resistively- and capacitively-shunted junction (RCSJ) model~\cite{Clarke}. In Fig.~\ref{Fig:Vphi}(b), four different regimes in the SQUID $V(\Phi)$ response are defined: $(i)$ the simple oscillatory regime where the bias current, $I_b=2.0I_0$. The other regimes are $(ii)$ the rapidly changing regime (red), $(iii)$ the zero voltage response regime (blue), and $(iv)$ the intermediate regime (green). Our analysis covers the interaction between a dc SQUID and an integrated cantilever when the system is tuned to operating points within these defined regimes, and the resulting effect on the cantilever-SQUID dynamics. 

\begin{figure}
\centerline{\includegraphics[width=3.4in,angle=0]{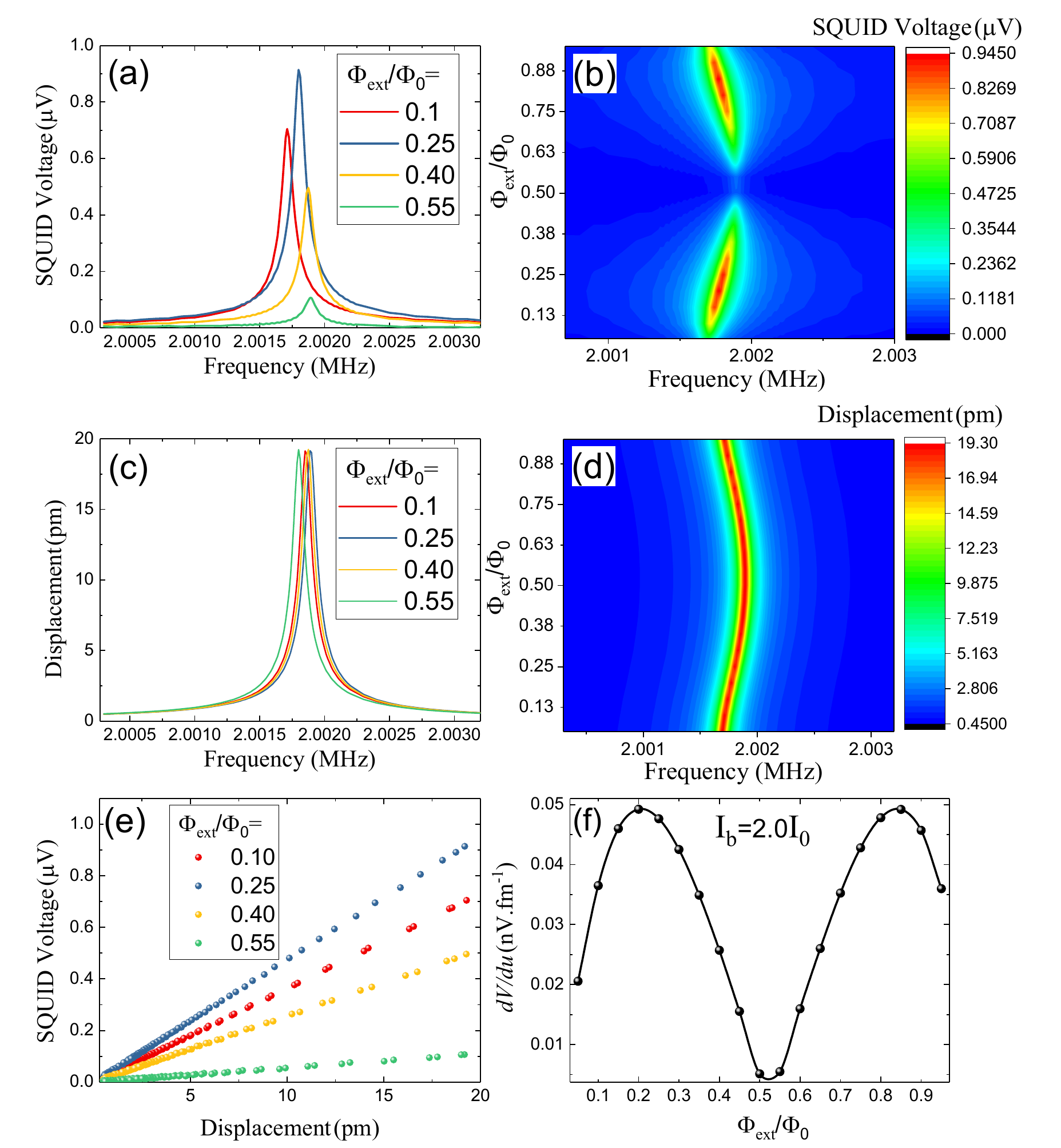}}
\caption{Line shapes for (a) SQUID voltage and (c)  cantilever displacement calculated as a function of $\Phi_{\mathrm{ext}}$ for $I_b=2.0I_0$. (b) and (d) density plots for SQUID voltages and cantilever displacement respectively. 
(e) the linear displacement-voltage trace as extracted by linking (b) and (d) via  the frequency. 
(f)  the  responsivity  ($\frac{\mathrm{d}V}{\mathrm{d}u}$) as calculated from the slopes of the displacement-voltage lines.}
%The  responsivity shows a sinusoidal behaviour which varies between  $4.7 \times 10^{-2}~\mathrm{nV.fm}^{-1}$ at $\Phi_{\mathrm{ext}}=0.25~\Phi_0$ and $0.5 \times 10^{-2}~\mathrm{nV.fm}^{-1}$ at $\Phi_{\mathrm{ext}}=0.50~\Phi_0$.}
\label{j2}
\end{figure}

We use the equation of motion of a damped harmonic oscillator given in~\cite{Poot} to describe the displacement, $u(t)$, of the mechanical cantilever:

\begin{equation}
m \ddot{u}+\frac{m \omega_0}{Q_0}\dot{u}+m \omega^2_0 u=F_d(t)+F_L(t),
\label{S1}
\end{equation}
where $m$ is the beam mass, $\omega_0=2 \pi f_0$ is the intrinsic frequency, $Q_0$ is the quality factor, $F_d=F_0 \cos(\omega_0 t)$ is the driving force, and $F_L(t)$  is  the Lorentz force $F_L(t)=aB\ell(I_b/2+J)$. Here, $B$ is the in-plane magnetic field, $\ell$ is the length of the cantilever,  $J$ is the circulating current, and $a= 0.91$~\cite{Etaki} is a geometrical factor that depends on the mode shape. Eq.~(\ref{S1}) is coupled to the dc SQUID equations given by the RCSJ model:

\begin{equation}
\frac{\Phi_0}{2\pi}C \ddot{\delta_1}+\frac{\Phi_0}{2\pi}C \frac{1}{R}\dot{\delta_1}+I_0 \sin(\delta_1)=\frac{1}{2} (I_b+J),
 \label{S2}
\end{equation}
 
\begin{equation}
\frac{\Phi_0}{2\pi}C \ddot{\delta_2}+\frac{\Phi_0}{2\pi}C \frac{1}{R}\dot{\delta_2}+I_0 \sin(\delta_2)=\frac{1}{2} (I_b-J),
 \label{S3}
\end{equation}

\begin{equation}
\delta_1- \delta_2=2\pi \cdot {\Phi_{\mathrm{tot}}}/{\Phi_0},
 \label{S4}
\end{equation}
where $\delta_{1,2}$ are the phase differences of the junctions, $\Phi_0$ is the flux quantum, $I_b$ is the bias current, $I_0$ is the critical current. The total flux, $\Phi_{\mathrm{tot}}$, has three contributions: $(i)$ the external flux $\Phi_{\mathrm{ext}}$, $(ii)$ the flux due to the circulating current, $J$, flowing through the inductance of the loop, $L$, and $(iii)$ the change in flux through the loop due to the cantilever displacement, $aB \ell u$. Therefore, $\Phi_{\mathrm{tot}}=\Phi_{\mathrm{ext}}+LJ + aB \ell u(t)$, and Eqs.~(\ref{S1}-\ref{S3}) are coupled via the circulating current  as $J=\frac{1}{L}(\frac{\delta_1-\delta_2}{2\pi}\Phi_0-\Phi_\mathrm{ext} - aB \ell u)$. 

These coupled differential equations are numerically solved without averaging the SQUID voltage and circulating currents, or scaling the time. Therefore, the time span $T_{\mathrm{max}}$ must be large enough to be suitable for the cantilever, while the time step $\mathrm{d}t$ must be small enough to resolve the impact of the fast changes dominated by the relatively high SQUID characteristic frequency $\omega_c=\frac{2\pi R I_0}{\Phi_0}$. Although this can be computationally expensive for cantilevers with very low frequencies relative to $\omega_c$, the experimental results of Etaki \textit{et al}~\cite{Etaki} allow their experiment to be modelled within a relatively small time window. 

Here, we solve a system for identical experimental conditions demonstrated by  Etaki \textit{et al} ~\cite{Etaki} with $f_0\simeq2$~MHz. To calculate the time dependent voltage, $V=\Phi_0 \frac{\dot{\delta_1}+\dot{\delta_2}}{2\pi}$, the Runge-Kutta method (RK4) was used to numerically integrate the equations presented above. The SQUID response was then obtained in the frequency domain by evaluating the Fourier transform of the SQUID voltage and the cantilever displacement. Our calculations were performed for $I_0=0.7$~$\mu$A, $R=29.5$~$\Omega$, $B=111$~mT, $C=0.91$~pF, and $L=170$~pH. These values give a McCumber-Stewart parameter $\beta_C= \frac{2 \pi I_0 R^2 C}{\Phi_0}=$ 1.61 and a screening parameter $\beta_L= \frac{2 I_0 L}{\Phi_0}=0.115$. The cantilever has a length $\ell=50$~$\mu$m, a mass $m=6 \times 10^{-13}$~kg, and was assumed to have a resonant frequency $f_0=2 .0018$~MHz and a quality factor $Q_0=25000$. The piezo drive which controls $F_\mathrm{d}$ is used only to locate the eigenmodes and is turned off during measurement~\cite{SEtaki}. Thus, at $t=0$ the initial velocity $v_0=\left.\frac{\mathrm du}{\mathrm dt}\right|_{u=u_0}=0$, 
where $u_0$ is the initial displacement amplitude. Here,  $u_0=20$~pm.

The time span chosen for these calculations was $T_{\mathrm{max}}= 25$~ms, i.e. more than six times the lifetime of the cantilever, and the optimised time step chosen was $\mathrm{d}t=0.0125$~ns. The calculations were repeated at different values of normalised flux in the range $0.90\Phi_0\leq \Phi_{\mathrm{ext}}\leq0.05\Phi_0$, and bias currents in 
the range $2.0I_0 \leq  I_\mathrm{b} \leq 1.10I_0$. In the frequency domain, we selected frequency steps of $\mathrm{d}f=12.5$~Hz. The units of the response which were calculated directly from a Fourier transform are V$\mathrm{\cdot }$s for the unnormalised SQUID voltage and $\mathrm{m.s}$ for the unnormalised cantilever displacement. To convert the units of the voltage-response from V$\mathrm{\cdot }$s to $\mathrm{V}$, the response was multiplied 
by $\frac{1}{\tau}$, where $\tau$ is the lifetime of the cantilever, which is related to the full width at half maximum (FWHM) as $\frac{1}{\pi \tau}=f_{\mathrm{FWHM}}$. A similar procedure was used to convert the units of the displacement-response from m$\mathrm{\cdot }$s to $\mathrm{m}$.

\section{\label{sec:level3}Results}
 \subsection{\label{sec:level3} The simple oscillatory regime behaviour}
To ensure our calculations are based in physical reality, we contextualized  the calculations with the existing experimental parameters.\cite{Etaki} As experimentally demonstrated, the voltage responses exhibit Lorentzian distributions and for $\Phi_\mathrm{ext}=0.25\Phi_0$, i.e. the highest SQUID sensitivity for $I_b=2.0I_0$ shown in Fig.~\ref{Fig:Vphi}(b), there was no relative experimental shift in $\Delta f$ of the cantilever. Changing the operating point of the SQUID by changing $\Phi_\mathrm{ext}$ within the simple oscillatory region shown in Fig. 1(b) affects $\Delta f$, and the operating point clearly affects the sensitivity to the SQUID voltage as clearly shown in Fig. 2(a) and (b). Moreover the subsequent cantilever displacement is also affected (Fig. 2(c) and (d)). These results clearly demonstrate the influence of the Lorentz back-action on the resonator from the SQUID displacement detector, and the expected magnitude of change in the experimental variables.

 \begin{figure}
\centerline{\includegraphics[width=3.7in,angle=0]{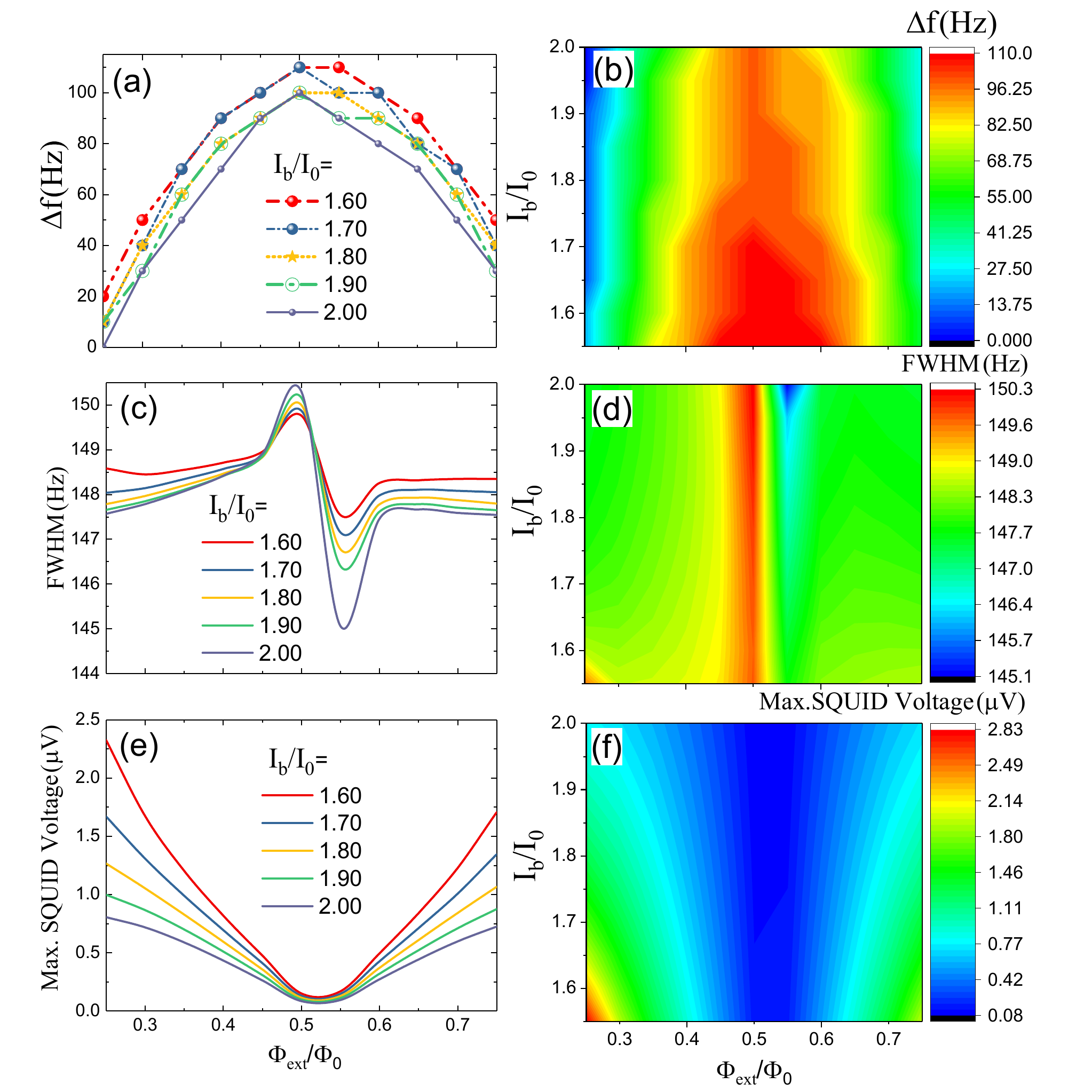}}
\caption{Calculations for the range $ 0.75\Phi_0\leq  \Phi_{\mathrm{ext}}\leq 0.25\Phi_0$ and $\mathrm{2.0I_0} \leq  I_\mathrm{b} \leq \mathrm{1.55I_0}$ for (a) the frequency shift, $\Delta f$, (c) FWHM and (e) the maximum SQUID voltage, $V_\mathrm{max}$.
The corresponding density plots are shown in (b), (d), and (f) respectively.}\label{j4}
\end{figure}

The cantilever displacement and SQUID voltage are explicitly linked in the frequency domain, i.e., the displacement $u(f)$ is parametrically linked to the voltage $V(f)$. The subsequent analysis was performed at $I_b=2.0I_0$ and $0.05\Phi_0 \leq \Phi_{\mathrm{ext}} \leq 0.95 \Phi_0$, and the displacement-voltage trace is plotted in Fig. 2(e). The traces show a linear dependence of voltage on displacement, which allows determination of the cantilever position in a responsivity specified by the slope of the displacement-voltage lines. Consequently, the responsivity ($\frac{\mathrm{d}V}{\mathrm{d}u}$) was calculated at $I_b=2.0I_0$ for different $\Phi_\mathrm{ext}$ values, with the result shown in Fig.~\ref{j2}(f). The figure shows a sinusoidal behaviour for $\frac{\mathrm{d}V}{\mathrm{d}u}$ which varies from $4.7 \times 10^{-2}~\mathrm{nV.fm}^{-1}$ at $\Phi_{\mathrm{ext}}=0.25\Phi_0$ 
to $0.5 \times 10^{-2}~\mathrm{nV.fm}^{-1}$ at $\Phi_{\mathrm{ext}}=0.50\Phi_0$. Importantly, Fig.~\ref{j2} shows an appropriate representation of the experimental results by Etaki~\textit{et al}~\cite{Etaki}, thereby demonstrating a good computational model.
 
 \subsection{\label{sec:level3} The intermediate regime behaviour} 
Further calculations were performed through the $V(\Phi)$ curve identified in Fig. 1(b) to examine the SQUID-cantilever coupling and explore the system response as the coupling/back-action is modified. Fig.~\ref{j4}(a)-(f) shows $\Delta f$, the FWHM, and the SQUID voltage as the bias current and $\Phi_{\mathrm{ext}}$ are tuned. The largest frequency shift corresponds to the smallest gradient ($\frac{\mathrm{d}V}{\mathrm{d}\Phi}$) of the working point. This can clearly be understood by Eq.~(1), where the frequency of the cantilever is controlled by the displacement coefficient. As the cantilever frequency is shifted by changing $\Phi_{\mathrm{ext}}$ and $I_b$, a modification in this coefficient emerges due to the circulating current dependence on $u$. Such a dependence was previously analysed by expanding the circulating current in terms of the displacement, $u$~\cite{Poot}. In this way, the new displacement coefficient, which arises from the back-action of the SQUID current on the cantilever, modifies the frequency and causes a slight or significant shift depending on $\Phi_{\mathrm{ext}}$ and $I_b$.

The Lorentz back-action also affects the cantilever quality factor; FWHMs of simulated line shapes are extracted and presented as a function of $\Phi_{\mathrm{ext}}$ for various values of $I_b$ in Fig.~\ref{j4}(c) and (d). The variation of the FWHM can be interpreted in an identical way to that of $\Delta f$, where the only difference being that $\mathrm{FWHM}=\frac{\omega_0}{2\pi Q_0 }$ is given in terms of velocity coefficient in Eq.~\ref{S1}. Thus, the FWHM is modified if $J$ is assumed to have a dependence on the velocity in addition to the displacement which modifies the frequency~\cite{Poot}. The corresponding peak voltage, $V_\mathrm{max}$, dependence on $\Phi_{\mathrm{ext}}$ and $I_b$ is shown in  Fig.~\ref{S3}(e) and (f). The behaviour of  $V_\mathrm{max}$ as a function of $\Phi_{\mathrm{ext}}$ is consistent with $\mathrm{d}V/\mathrm{d}\Phi_{\mathrm{ext}}$ of the SQUID $V(\Phi_{\mathrm{ext}})$ curve shown in Fig.~\ref{Fig:Vphi}(b).

\begin{figure}
\centerline{\includegraphics[width=3.0in,angle=0]{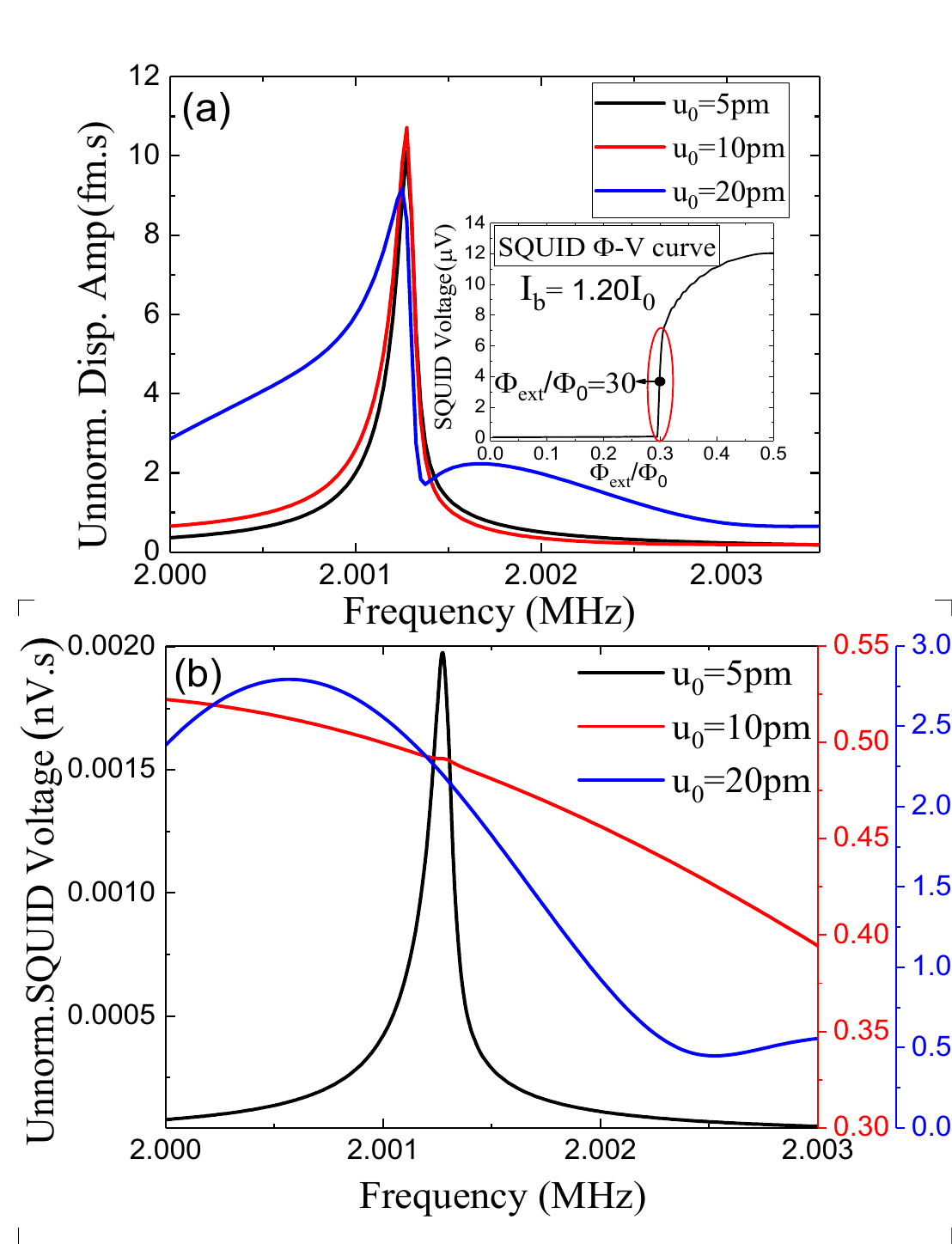}}
\caption{(a) unnormalised displacement and (b) corresponding unnormalised SQUID voltage when the SQUID displacement detector is tuned ($\Phi_{\mathrm{ext}}=0.30\Phi_0$ and $I_b=1.20I_0$) to the working point shown in the inset of (a). The initial cantilever amplitudes are $u_0=20 ~\mathrm{pm}$ (blue), $u_0=10~ \mathrm{pm}$ (red), and $u_0=5 ~\mathrm{pm}$ (black), which correspond to a change of flux in the SQUID loop of $0.05\Phi_0$, $0.025\Phi_0$, and $0.0125\Phi_0$ respectively.}
\end{figure}

\begin{figure}
\centerline{\includegraphics[width=3.6in,angle=0]{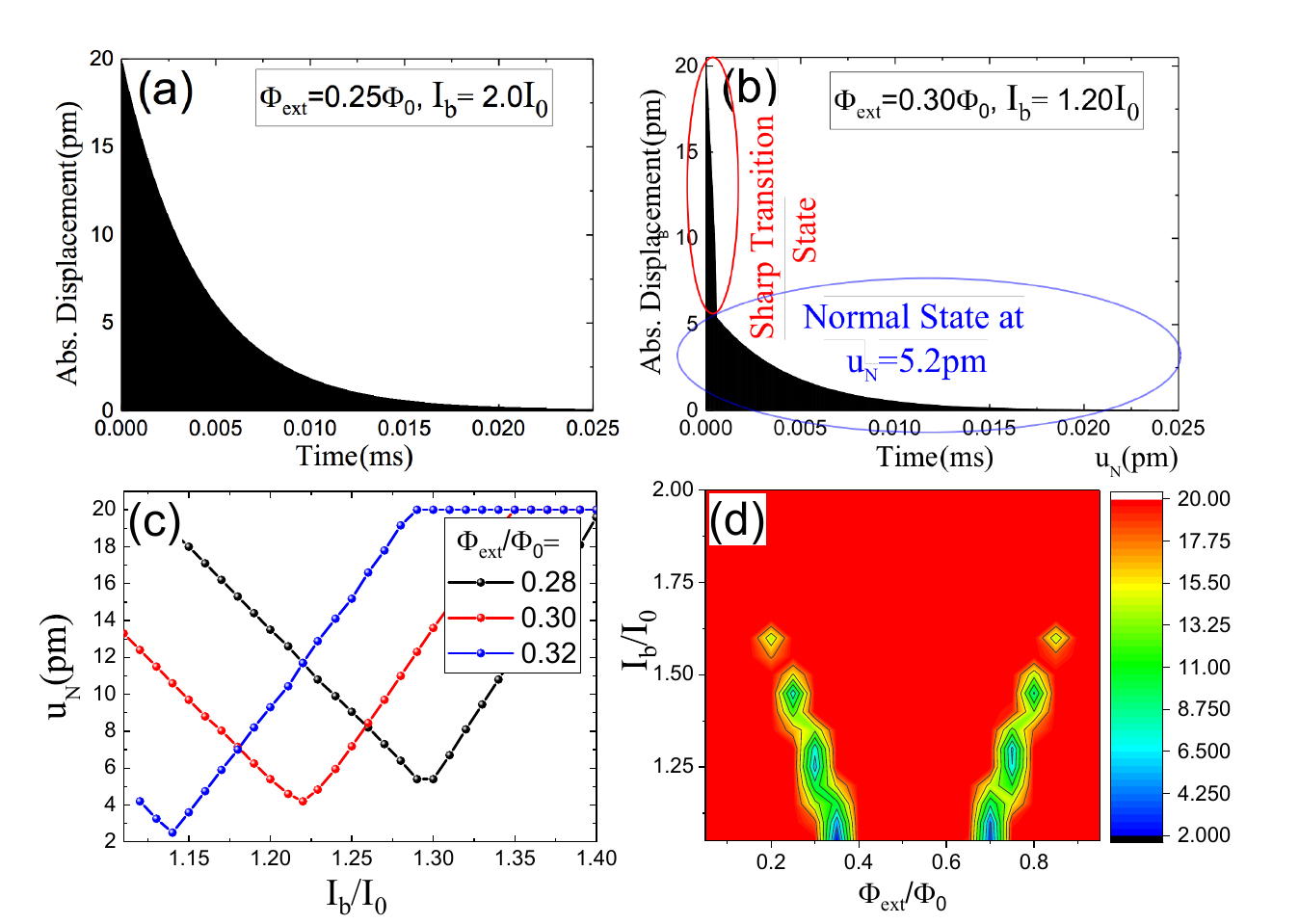}}
\caption{ 
(a) Snapshot for the time array of cantilever displacements at (a) $\Phi_{\mathrm{ext}}=0.25\Phi_0$ and $I_b=2.0I_0$ (a point in the simple oscillatory  regime) versus (b)  a point $\Phi_{\mathrm{ext}}=0.30\Phi_0$ and $I_b=1.20I_0$ in the rapidly changing regime
in which a sharp transition state emerges until the cantilever enters  the normal state at $u=u_\mathrm{N}$. (c) calculations for the normal state  positions of specific lines around $\Phi_{\mathrm{ext}}=0.30\Phi_0$ and $I_b=1.20I_0$, and (d) the yellow-blue islands in the density plot indicate 
 a  shift in the normal state positions that starts emerging at $t=t_\mathrm{N}$ and $u=u_\mathrm{N}$.}
\end{figure}

 \subsection{\label{sec:level3} The rapidly changing regime behaviour} 
Now we turn to a different regime from Fig 1(b), where the largest effect of back-action on the cantilever is observed, and the SQUID response is apparently nonlinear. To examine the effect of back-action on the cantilever motion, a point in such region was selected as shown in the inset of Fig. 4(a). Subsequently, at $\Phi_{\mathrm{ext}}=0.30\Phi_0$ and $I_b=1.20I_0$, the unnormalised cantilever displacement and corresponding unnormalised SQUID voltage response for various displacement amplitudes $u_0$, are obtained and plotted in Fig. 4(a) and Fig. 4(b). When  $u_0=20 ~\mathrm{pm}$, the cantilever appears to have a nonlinear behaviour as demonstrated by the modified  line shape of the cantilever and the SQUID response. Specifically, as the displacement is reduced from $u_0=20~\mathrm{pm}$ to $u_0=10 ~\mathrm{pm}$ to $u_0=5 \mathrm{pm}$, the change in the flux through the loop is $0.05$, $0.025$ and $0.0125\Phi_0$ respectively. This change of flux in the SQUID loops drives the cantilever to experience two $V(\Phi)$ regions of different responsivity, which results in the nonlinear-like behaviour. It should be noted, however, that as the cantilever returns to its dynamical equilibrium position, the response becomes more Lorentzian as expected.

The effect of the SQUID-cantilever interaction on the cantilever motion can be more clearly observed by comparing the time evolution of the cantilever displacement for two different bias and flux values. The time dependent displacement for $\Phi_{\mathrm{ext}}=0.25\Phi_0$ and  $I_b=2.0I_0$ is plotted in Fig. 5(a), and for $\Phi_{\mathrm{ext}}=0.30\Phi_0$ and  $I_b=1.20I_0$ in Fig. 5(b). Clearly if the SQUID operating point is in the rapidly changing regime (Fig. 5b) there is a sharp transition state as the cantilever returns to its equilibrium position. Naively if the SQUID bias is switched when the cantilever motion is large, there is an instantaneous damping which can be used to modify the motion of the cantilever. Normal state positions, $u_\mathrm{N}$, for specific lines around $\Phi_{\mathrm{ext}}=0.30\Phi_0$ and  $I_b=1.20I_0$ are shown in Fig. 5(c). These positions are extracted when the cantilever enters the normal state that accounts for the Lorentzian profile in the frequency domain, and when the amplitude starts decaying exponentially at time $t=t_\mathrm{N}$, as shown in  Fig. 5(b). A more comprehensive analysis is presented in the density plot shown in Fig. 5(d). The plot given in Fig. 5(c) exhibits details for one of the yellow-blue islands in the density plot. The islands correspond to the intermediate regimes in the $V(\Phi)$ curves.
Its anticipated that such effect could be employed to precisely and rapidly control the amplitude of the cantilever displacement below its initial amplitude which can be set by a piezo drive used to locate the eigenmodes of the cantilever. In other words, putting the system in such regions  enables
modulating the cantilever amplitude after isolating the system from the external actuator. 

\section{\label{sec:level3}Conclusion}
In conclusion, we have shown how the tuning of the SQUID device affects the back-action between the SQUID and the doubly clamped cantilever. Specifically, we have quantified the line shapes expected from the SQUID response and the corresponding cantilever displacement. The effect can be quantitatively analysed via the shift in the cantilever frequency, the line width, intensity, and shift in the position of the normal state. Direct solutions for the unscaled dc SQUID equations coupled to the equations of motion of an integrated cantilever allow determination of voltage-displacement traces of
a displacement detector. For a SQUID displacement detector tuned to a working point in the rapidly changing region, a sharp transition state emerges and a nonlinear-like response due to the emergence of such state is observed. This state could be used to employ the system as a self modulator for the displacement amplitude of the cantilever.  
These results should allow a clearer understanding and manipulation of future experimental work.

\begin{acknowledgments}
The authors would like to acknowledge Andrew Armour for numerous conversations and beneficial comments and N. Peretto, R. M. Smith, and R. A. Frewin for the computational facilities, and E. Riordan for reading and discussions. We also gratefully acknowledge support by the European Research Council under the EU Consolidator Grant "SUPERNEMS" (Project No. 647471). 
\end{acknowledgments}

\nocite{*}
\bibliography{aipsamp}% Produces the bibliography via BibTeX.

\end{document}